\documentclass[10pt,twocolumn]{article}
\usepackage[top=1.8cm, bottom=1.8cm, left=1.8cm, right=1.8cm]{geometry}
\usepackage{times} 

\usepackage{titlesec} 
\titlespacing*{\section}{0pt}{*3}{3pt} 
\titlespacing{\subsection}{0pt}{*2}{2pt}

\usepackage[hyphens]{url} 
\usepackage{graphicx}
\def\UrlFont{\rm} 
\usepackage{graphicx} 
\usepackage{url}                                %
\usepackage{array,multirow,graphicx,adjustbox}  %
\usepackage{booktabs}                           %
\usepackage[utf8]{inputenc}
\usepackage[ruled,algosection,noend,linesnumbered]{algorithm2e}
\usepackage{float}
\usepackage{paralist}
\usepackage{amsmath}
\usepackage{amsfonts}
\usepackage{xcolor}
\usepackage{csquotes} 
\usepackage{tabularx}
\usepackage{makecell}
\usepackage{pbox}
\usepackage[keeplastbox]{flushend}
\usepackage{subfigure}
\usepackage[small,bf]{caption}
\usepackage[hang,flushmargin]{footmisc}
\usepackage{xspace}
\usepackage{multicol, lipsum}
\usepackage[T1]{fontenc}

\usepackage[bookmarks=true, bookmarksnumbered=true, colorlinks=true, linkcolor=linkcol, citecolor=citecol, urlcolor=urlcol, hypertexnames=true]{hyperref}

\definecolor{darkgreen}{RGB}{0, 100, 0}
\definecolor{linkcol}{rgb}{0.3,0,0}
\definecolor{citecol}{rgb}{0.3,0,0}
\definecolor{urlcol}{rgb}{0.3,0,0}

\makeatletter
\def\url@leostyle{%
  \@ifundefined{selectfont}{\def\UrlFont{\small}}%
  {\def\UrlFont{}}%
}
\makeatother
\newcommand{\descr}[1]{\smallskip\noindent\textbf{#1}}

\let\OLDthebibliography\thebibliography
\renewcommand\thebibliography[1]{
  \OLDthebibliography{#1}
  \setlength{\parskip}{0pt}
  \setlength{\itemsep}{1pt plus 0.2ex}
}

\newif
\ifcomment
\commentfalse
\ifcomment
\newcommand{\sz}[1]{{\bf \textcolor{brown}{SZ: #1}}}
\newcommand{\edc}[1]{{\bf \textcolor{red}{EDC: #1}}}
\newcommand{\gs}[1]{{\bf \textcolor{orange}{GS: #1}}}
\newcommand{\ap}[1]{{\bf \textcolor{blue}{AP: #1}}}
\newcommand{\jbnote}[1]{{\bf \textcolor{magenta}{JB: #1}}}
\newcommand{\ma}[1]{{\bf \textcolor{pink} {MAX: #1}}}
\else
\newcommand{\sz}[1]{}
\newcommand{\edc}[1]{}
\newcommand{\gs}[1]{}
\newcommand{\ap}[1]{}
\newcommand{\jbnote}[1]{}
\newcommand{\ma}[1]{}
\fi

\title{\bf An Early Look at the Parler Online Social Network} 

\author{Max Aliapoulios$^1$, Emmi Bevensee$^2$, Jeremy Blackburn$^3$, Barry Bradlyn$^4$,\\Emiliano De Cristofaro$^5$, Gianluca Stringhini$^6$, and Savvas Zannettou$^7$    \\[0.7ex]
\normalsize{$^{1}$New York University, $^2$SMAT, $^3$Binghamton University, $^4$University of Illinois at Urbana-Champaign,}\\
\normalsize{$^5$University College London, $^6$Boston University, $^7$Max Planck Institute for Informatics}\\
\normalsize maliapoulios@nyu.edu, rebelliousdata@gmail.com, jblackbu@binghamton.edu, bbradlyn@illinois.edu, e.decristofaro@ucl.ac.uk\\ \normalsize gian@bu.edu, szannett@mpi-inf.mpg.de\vspace*{-0.2cm}}

\date{}

\begin{document}
\maketitle

\begin{abstract}
Parler is as an ``alternative'' social network promoting itself as a service that allows to ``speak freely and express yourself openly, without fear of being deplatformed for your views.''
Because of this promise, the platform become popular among users who were suspended on mainstream social networks for violating their terms of service, as well as those fearing censorship.
In particular, the service was endorsed by several conservative public figures, encouraging people to migrate from traditional social networks.
After the storming of the US Capitol on January 6, 2021, Parler has been progressively deplatformed, as its app was removed from Apple/Google Play stores and the website taken down by the hosting provider.

This paper presents a dataset of 183M Parler posts made by 4M users between August 2018 and January 2021, as well as metadata from 13.25M user profiles.
We also present a basic characterization of the dataset, which shows that the platform has witnessed large influxes of new users after being endorsed by popular figures, as well as a reaction to the 2020 US Presidential Election.
We also show that discussion on the platform is dominated by conservative topics, President Trump, as well as conspiracy theories like QAnon.
\end{abstract}

\section{Introduction}

Over the past few years, social media platforms that cater specifically to users disaffected by the policies of mainstream social networks have emerged.
Typically, these tend not to be terribly innovative in terms of features, but instead attract users based on their commitment to ``free speech.''
In reality, these platforms usually wind up as echo chambers, harboring dangerous conspiracies and violent extremist groups.
A case in point is Gab, one of the earliest alternative homes for people banned from Twitter~\cite{zannettou2018gab,fair2019shouting}. %
After the Tree of Life terrorist attack, it %
was hit with multiple attempts to de-platform the service, essentially erasing it from the Web.
Gab, however, has survived and even rolled out new features under the guise of free speech that are in reality tools used to further evade and circumvent moderation policies put in place by mainstream platforms~\cite{dissenterimc}.

Worryingly, Gab as well as other more fringe platforms like Voat~\cite{papasavva2020qoincidence} and TheDonald.win~\cite{manoel_bans} have shown that not only is it feasible in technical terms to create a new social media platform, but marketing the platform towards specific polarized communities is an extremely successful strategy to bootstrap a user base.
In other words, there is a subset of users on Twitter, Facebook, Reddit, etc., that will happily migrate to a new platform, especially if it advertises moderation policies that do not restrict the growth and spread of political polarization, conspiracy theories, extremist ideology, hateful and violent speech, and mis- and dis-information.

\descr{Parler.} In this paper, we present an extensive dataset collected from Parler.
Parler is an emerging social media platform that has positioned itself as the new home of disaffected right-wing social media users in the wake of active measures by mainstream platforms to excise themselves of dangerous communities and content.
While Parler works approximately the same as Twitter and Gab, it additionally offers an extensive set of self-serve moderation tools.
For instance, filters can be set to place replies to posted content into a moderation queue requiring manual approval, mark content as spam, and even automatically block all interactions with users that post content matching the filters. %

After the events of January 6, 2021, when a violent mob stormed the US capitol, Parler came under fire for letting threat of violence unchallenged on its platform.
The Parler app was first removed from the Google Play and the Apple App stores, and the website was eventually deplatformed by the hosting provider, Amazon AWS.
At the time of writing, it is unclear whether Parler will come back online and when.

\descr{Data Release.} Along with the paper, we release a dataset~\cite{zenodo} including 183M posts made by 4M users between August 2018 and January 2021, as well as metadata from 13.25M user profiles.
Each post in our dataset has the content of the post along with other metadata (creation timestamp, score, hashtags, etc.).
The profile metadata include bio, number of followers, how many posts the account made, etc.
Our data release follows the FAIR principles, as discussed later in Section~\ref{sec:data_structure}.

We warn the readers that we post and analyze the dataset unfiltered; as such, some of the content might be toxic, racist, and hateful, and can overall be disturbing.

\descr{Relevance.} We are confident that our dataset will be useful to the research community in several ways. 
Parler gained quick popularity at a very crucial time in US History, following the refusal of a sitting President to concede a lost election, an insurrection where a mob stormed the US Capitol building, and, perhaps more importantly, the unprecedented ban of the US President platforms like Facebook and Twitter.
Thus, the dataset will constitute an invaluable resource for researchers, journalists, and activists alike to study this particular moment following a data-driven approach.

Moreover, Parler attracted a large migration of users on the basis of fighting censorship, reacting to deplatforming from mainstream social networks, and overall an ideology of striving toward unrestricted online free speech.
As such, this dataset provides an almost unique view into the effects of deplatforming as well as the rise of a social network specifically targeted to a certain type of users.
Finally, our Parler dataset contains a large amount of hate speech and coded language that can be leveraged to establish baseline comparisons as well as to train classifiers.

\section{What is Parler?}
\label{sec:background}

Parler (usually pronounced ``par-luh'' as in the French word for ``to speak'') is a microblogging social network launched in August 2018.
Parler markets itself as being ``built upon a foundation of respect for privacy and personal data, free speech, free markets, and ethical, transparent corporate policy''~\cite{syracuse}.
Overall, Parler has been extensively covered in the news for fostering a substantial user-base of Donald Trump supporters, conservatives, conspiracy theorists, and right-wing extremists~\cite{wapo2020parler}. %

\descr{Basics.} At the time of our data collection, to create an account, users had to provide an email address and phone number that can receive an activation SMS (Google Voice/VoIP numbers are not allowed).
Users interact on the social network by making posts of maximum 1,000 characters, called ``parlays,'' which are broadcasted to their followers.
Users also have the ability to make comments on posts and on other comments.

\descr{Voting.}
Similar to Reddit and Gab, Parler also has a voting system designated for ranking content, following a simple upvote/downvote mechanism.
Posts can only be upvoted, thus making upvotes functionally similar to likes on Facebook.
Comments to posts, however, can receive both upvotes and downvotes.
Voting allows users to influence the order in which comments are displayed, akin to Reddit score.

\descr{Verification.}
Verification on Parler is opt-in; %
users can willingly make a verification request by submitting a photograph of themselves and a photo-id card. 
According to the website, verification--in addition to giving users a red badge--evidently ``unlocks additional features and privileges.'' 
They also declare that the personal information required for verification is never shared with third parties, and that after verification such information is deleted except for ``encrypted selfie data.'' 
At the time of writing, only 240,666 (2\%) users on Parler are verified.

\descr{Moderation.}
The Parler platform  has the capability to perform content moderation and user banning through administrators.
We explore these functionalities, from a quantitative perspective, in Section~\ref{sec:user_analysis}. 
Note that there are several moderation attributes put in place per account, which are visible in an account's settings. 
For instance, there is a field for whether the account ``pending.''
It appears that new accounts show up as ``pending'' until they are approved by automated moderation. 

Each account has a ``moderation'' panel allowing users to view comments on their own content and perform moderation actions on them. 
A comment can fall into any of five moderation categories: {\em review}, {\em approved}, {\em denied}, {\em spam}, or {\em muted}. 
Users can also apply keyword filters, which will enact one of several automated actions based on a filter match: {\em default} (prevent the comment), {\em approve} (require user approval), {\em pending}, {\em ban member notification}, {\em deny}, {\em deny with notification}, {\em deny detailed}, {\em mute comment}, {\em mute member}, {\em none}, {\em review}, and {\em temporary ban}. 
These actions are enforced at the level of the user configuring the filters, i.e., if a filter is matched for temporary ban, then the user making the comment matching the filter is banned from commenting on the original user's content.

There are several additional comment moderation settings available to users. 
For example, users can allow only verified users to comment on their content; %
there are tools to handle spam, etc. %
Overall, Parler allows for more individual content moderation compared to other social networks; however, recent reports have highlighted how global moderation is arguably weaker, as large amounts of illegal content has been allowed on the platform~\cite{wapo2020parlerporn}. 
We posit this may be due to global moderation being a {\em manual} process performed by a few accounts. %

\descr{Monetization.}
Parler supports ``tipping,'' allowing users to tip one another for content they produce. 
This behavior is turned off by default, both with respect to accepting and being able to give out tips. 
An additional monetization layer is incorporated within Parler, which is called ``Ad Network'' or ``Influence Network''~\cite{donk_enby_2021_4426480}.
Users with access to this feature are able to pay for or earn money for hosting ad campaigns. 
Users set their rate per thousand views in a Parler specified currency called ``Parler Influence Credit.''

\section{Data Collection}
We now discuss our methodology to build the dataset released along with this paper.
We use a custom-built crawler that accesses the (undocumented, but open) Parler API.
This crawler was based on Parler API discoveries that allowed for faster crawling~\cite{donk_enby_2020_4426283}.

\descr{Crawling.} Our data collection procedure works as follows.
First, we populate users via an API request that maps a monotonically increasing integer ID (modulo a few exceptions) %
to a universally unique ID (UUID) that serves as the user's ID in the rest of the API. %
Next, for each UUID we discover, we query for its profile information, which includes metadata such as badges, whether or not the user is banned, bio, public posts, comments, follower and following counts, when the user joined, the user's name, their username, whether or not the account is private, whether or not they are verified, etc. 
Note that, to retrieve posts and comments, we use an API endpoint that allows for time-bounded queries; i.e., for each user, we retrieve the set of post/comments since the most recent post/comment we have already collected for that user.

\descr{Data.} Overall, we collect all user profile information for the 13.25M Parler accounts created between August 2018 and January 2021.
Additionally, we collect 98.5M public posts and 84.5M public comments from a random set of 4M users; see Table~\ref{tab:dataset}.
As mentioned, the dataset is available from~\cite{zenodo}.

\descr{Limitations of Sampling.} As mentioned above, posts and comments in our dataset are from a sample of users; more precisely, 183.063M and 4.08M users, respectively.
Although, numerically, this should in theory provide us with a good representation of the activities of Parler's user base, we acknowledge that our sampling might {\em not} necessarily be representative in a strict statistical sense.
In fact, using a two-sample KS test, we reject the null hypothesis that the distribution of comments reported in the profile data from all users is the same as the distribution of those we actually collect from 1.1M users ($p < 0.01$).
We speculate that this could be due to the presence of a small number of very active users which were not captured in our sample.
Moreover, we have posts from many fewer users than we have comments for; this is due to users' tendency to make more posts than comments, which increases the wall clock time it takes to collect posts.

Therefore, we need to take these possible limitations into account when analyzing user content---e.g., as we do in Section~\ref{sec:content_analysis}.
Nonetheless, we believe that our sample does ultimately capture the general trends measured from profile data, and thus we are confident our sample provides at least a reasonable representation of content posted to Parler.

\descr{Ethical Considerations.} We only collect and analyze publicly available data. 
We also follow standard ethical guidelines~\cite{rivers2014ethical}, not making any attempts to track users across sites or de-anonymize them. 
Also, taking into account user privacy, we remove from the data the names of the Parler accounts in our dataset.

\begin{table}[t]
  \centering
  \resizebox{\columnwidth}{!}{%
  \begin{tabular}{@{}lrrrr@{}}
  \toprule
  \textbf{}            & \multicolumn{1}{c}{\textbf{Count}} & \multicolumn{1}{c}{\textbf{\#Users}} & \multicolumn{1}{c}{\textbf{Min. Date}} & \multicolumn{1}{c}{\textbf{Max. Date}} \\ \midrule
  \textbf{Posts} & 98,509,761                            & 2,439,546                               & 2018-08-01                             & 2021-01-11                             \\
  \textbf{Comments}    & 84,546,856                           & 2,396,530                              & 2018-08-24                             & 2021-01-11                             \\ \midrule
  \textbf{Total}       & 183,056,617                           & 4,079,765                              & 2018-08-01                             & 2021-01-11                    \\ \bottomrule
  \end{tabular}%
  }
  \caption{Dataset Statistics.}
  \label{tab:dataset}
  \end{table}

\section{Data Structure}\label{sec:data_structure}

This section presents the structure of the data, available at~\cite{zenodo}.
Overall, the data consists of newline-delimited JSON files (\texttt{.ndjson}), obtained by crawling three main Parler API endpoints, \texttt{/v1/post}, \texttt{/v1/comment}, and \texttt{/v1/user}.
Each JSON consists of key/value pairs returned by their respective API endpoint.

Due to space limitations, in the following, we only list the keys used in the analysis in this paper.
The complete key/value list as well as their definitions are available at~\cite{zenodo}.

\descr{Key/Values from /v1/post[comment].}
From the post and comment endpoints, we have:
\begin{compactitem}
  \item {\em id}: Parler generated UUID of the post/comment. 
  \item {\em createdAt}: Timestamp of the post/comment in UTC. 
  \item {\em upvotes}: Number of upvotes that a post/comment received.
  \item {\em score}: Number of upvotes minus the sum of the downvotes a post/comment received.
  \item {\em hashtags}: List of strings that corresponds to the hashtags used in a post/comment.
  \item {\em urls}: List of dictionaries correspond to URLs and their respective metadata used in a post/comment.
\end{compactitem}

We also enriched the data with fields from the user profile who produced the content, like \textit{username} and \textit{verified}, in order to provide additional context to the post or comment. 
Additionally, we formatted the following fields from strings to integers to facilitate numerical analysis: \textit{depth}, \textit{impressions}, \textit{reposts}, \textit{upvotes}, and \textit{score}.

\descr{Key/Values from \texttt{/v1/user}.} From the user endpoint, we have:
\begin{compactitem}
  \item {\em id}: Parler generated UUID for the user.
  \item {\em badges}: List of numeric values corresponding to the user profile badges.
  \item {\em bio}: Biography string written by a user for their profile.
  \item {\em ban}: Boolean field as to whether or not the user is currently banned.
  \item {\em user\_followers}: Numeric field corresponding to the number of followers the user profile has.
  \item {\em user\_following}: Numeric field corresponding to the number of users the user profile follows.
  \item {\em posts}: Number of posts a user has made.
  \item {\em comments}: Number of comments a user has made.
  \item {\em joined}: Timestamp of when a user joined in UTC.
\end{compactitem}

We renamed the following fields because they are reserved field names in our datastore: \textit{followers} to \textit{user\_followers} and \textit{following} to \textit{user\_following}. 
We also reformatted the following fields from strings to integers to facilitate numerical analysis: \textit{comments}, \textit{posts}, \textit{following}, \textit{media}, \textit{score}, \textit{followers}.

\descr{FAIR Principles.} The data released along with this paper aligns with the FAIR guiding principles for scientific data.\footnote{\url{https://www.go-fair.org/fair-principles/}} 
First, we make our data \emph{Findable} by assigning a unique and persistent digital object identifier (DOI): 10.5281/zenodo.4442460.
Second, our dataset is \emph{Accessible} as it can be downloaded, for free.
It is in JSON format, which is widely used for storing data and has an extensive and detailed documentation for all of the computer programming languages that support it, thus enabling our data to be \emph{Interoperable}.
Finally, our dataset is extensively documented and described in this paper and in~\cite{zenodo}, and released openly, thus our dataset is \emph{Reusable}.

\section{User Analysis}
\label{sec:user_analysis}
This section analyzes the data from the 13.25M Parler user profiles collected between November 25, 2020 and January 11, 2021.

\subsection{User Bios}
We analyze the user bios of all the Parler users in our dataset.
We extract the most popular words and bigrams, reported in Table~\ref{table:user_profile_table}.
Several popular words indicate that a substantial number of users on Parler self identify as conservatives (1.3\% of all users include the word ``conservative'' it in their bios), Trump supporters (1\% include the word ``trump'' and 0.27\% the bigram ``trump supporter''), patriots (0.79\% of all users include the word ``patriot''), and religious individuals (1.05\% of all users include the word ``god'' in their bios).
Overall, these results indicate that Parler attracts a user base similar to the one that exists on Gab~\cite{zannettou2018gab}.

\begin{table}[t]
  \centering
  \small
  \begin{tabular}{lr|lr}
  \toprule
    \textbf{Word}         &   \multicolumn{1}{r}{\bf (\%)}    &         \textbf{Bigram}  &  \textbf{(\%)}     \\
    \midrule
    conservative &  1.23\% &         trump supporter &  0.26\% \\
    god          &  0.99\% &          husband father &  0.24\% \\
    trump        &  0.96\% &             wife mother &  0.19\% \\
    love         &  0.88\% &              god family &  0.17\% \\
    christian    &  0.78\% &              trump 2020 &  0.17\% \\
    patriot      &  0.76\% &          proud american &  0.17\% \\
    wife         &  0.74\% &                wife mom &  0.16\% \\
    american     &   0.7\% &                pro life &  0.15\% \\
    country      &  0.65\% &  christian conservative &  0.14\% \\
    family       &  0.62\% &            love country &  0.13\% \\
    life         &  0.58\% &                love god &  0.13\% \\
    proud        &  0.57\% &          family country &  0.13\% \\
    maga         &  0.55\% &         president trump &  0.12\% \\
    mom          &  0.54\% &               god bless &  0.12\% \\
    father       &  0.54\% &          business owner &  0.12\% \\
    husband      &  0.52\% &            jesus christ &   0.1\% \\
    jesus        &  0.45\% &  conservative christian &   0.1\% \\
    freedom      &  0.43\% &        american patriot &   0.1\% \\
    retired      &  0.42\% &                maga kag &   0.1\% \\
    america      &  0.41\% &             god country &  0.09\% \\
    \bottomrule
    \end{tabular}
 \caption{Top 20 words and bigrams found in Parler users bios.}
 \label{table:user_profile_table}
 \end{table}

   \begin{table}[t]
    \small
\setlength{\tabcolsep}{2pt}
        \centering
        \begin{tabular}{@{}rrp{1.375cm}p{5cm}@{}}
          \toprule
          \textbf{Badge} & \textbf{\#Unique} & \textbf{Badge}   & \textbf{Description}                                                                                                                                                                                                                           \\ 
          {\bf \#} & {\bf Users} & {\bf Tag} \\      \midrule
          0                 & 250,796                  & Verified             & Users who have gone through verification.                                                                                                                                                                                                      \\ \hline
          1                 & 605                      & Gold                 & Users whom Parler claims may attract targeting, impersonation, or phishing campaigns.                                                                                                                                      \\ \hline
          2                 & 81                       & Integration Partner  & Used by publishers to import all their articles, content, and comments from their website.                                                                                                                                                  \\ \hline
          3                 & 112                      & \begin{tabular}[t]{@{}l}Affiliate\\(RSS Feed)\end{tabular}  & Shown as affiliates in the website but known as RSS feed to Parler mobile apps. Integrated directly into an existing off-platform feed. Share content on update to an RSS feed. \\ \hline
          4                 & 922,695                  & Private              & Users with private accounts.                                                                                                                                                                                                                   \\ \hline
          5                 & 4,908                    & Verified Comments    & User is verified (badge 0) and is restricting comments only to other verified users.                                                                                                                                      \\ \hline
          6                 & 49                       & Parody               & Users with ``approved'' parody. Despite guidelines against impersonation, some are allowed if approved as parody.                                                          \\ \hline
          7                 & 34                       & Employee             & Parler employee.                                                                                                                                                                                                         \\ \hline
          8                 & 2                        & Real Name            & Using real name as their display name. Not clear how/if this information is verified.                                                                                                                                     \\ \hline
          9                 & 1030                      & Early Parley-er      & Joined Parler, and was active, early on (on or before December 30th, 2018).       \\ \bottomrule
          \end{tabular}
        \caption{Badges assigned to user profiles. Users are given an array of badges to choose from based on their profile parameters.}
        \label{tab:badges}
        \end{table}

\subsection{Bans}
Parler profile data includes a flag that is set when a user is banned. 
We find this flag set for 252,209 (2.09\%) of users.
Almost all of these banned accounts, 252,076 (99.95\%) are also set to private, but for those that are not, we can observe their username, name and bio attributes, and even retrieve comments/posts they might have made.
While not a thorough analysis of the ban system, when exploring the 157 non-private banned accounts, we notice some interesting things.
In general, there appears to be two classes of banned users.
The first are accounts banned for impersonating notable figures, e.g., the name ``Donald J Trump'' and a bio that describes the user as the ``45th President of the United States of America,'' or a ``ParlerCEO'' username with ``John Matza'' as the user's display name.
These actions violate the %
Parler guideline around ``Fraud, IP Theft, Impersonation, Doxxing'' suggesting Parler does in fact enforce at least some of their moderation policies.
For the second class of banned user, it is harder to determine the guideline violation that led to the ban.
For example, an account named ``ConservativesAreRetarded'' whose profile picture was a hammer and sickle made a comment (the account's only comment) in response to a post by a Parler employee.
The comment,``You look like garbage, at least take a decent photo of the shirt,'' was made in reply to a post that included an image the Parler employee wearing a Parler t-shirt.
While the comment by ``ConservativesAreRetarded'' is certainly not nice, it was not clear to us which of the Parler guidelines it violated.
We do not believe this would be considered violent, threatening, or sexual content, which are explicitly noted in the guidelines.
Although outside the scope of this paper, it does call into question how consistently Parler's moderation guidelines are followed.

\subsection{Badges}

There are several badges that can be awarded to a Parler user profile. These badges correspond to different types of account behavior. Users are able to select which badges they opt to appear on their user profile.%
We detail the large variety of badges available in Table~\ref{tab:badges}.
A user can have no badges or multiple badges. 
For each user, our crawler returns a set of badge numbers; we then looked up users with specific badges in the Parler UI in order to see the badge tag and description which are displayed visually.

\begin{figure*}[t!]
  \centering
  \subfigure[]{\includegraphics[width=0.8\columnwidth]{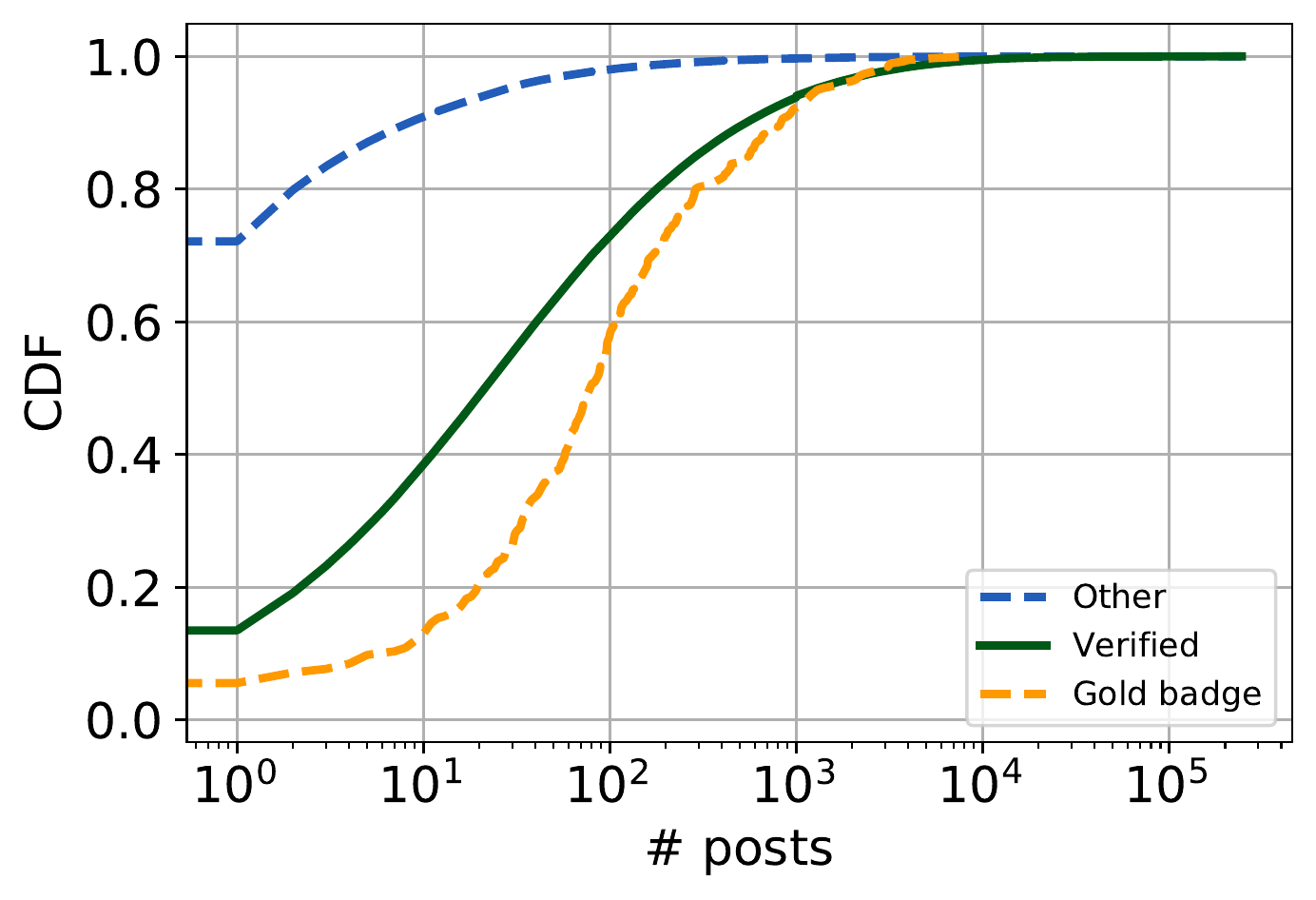}\label{fig:verified_posts_cdf}}
~~~
    \subfigure[]{\includegraphics[width=0.8\columnwidth]{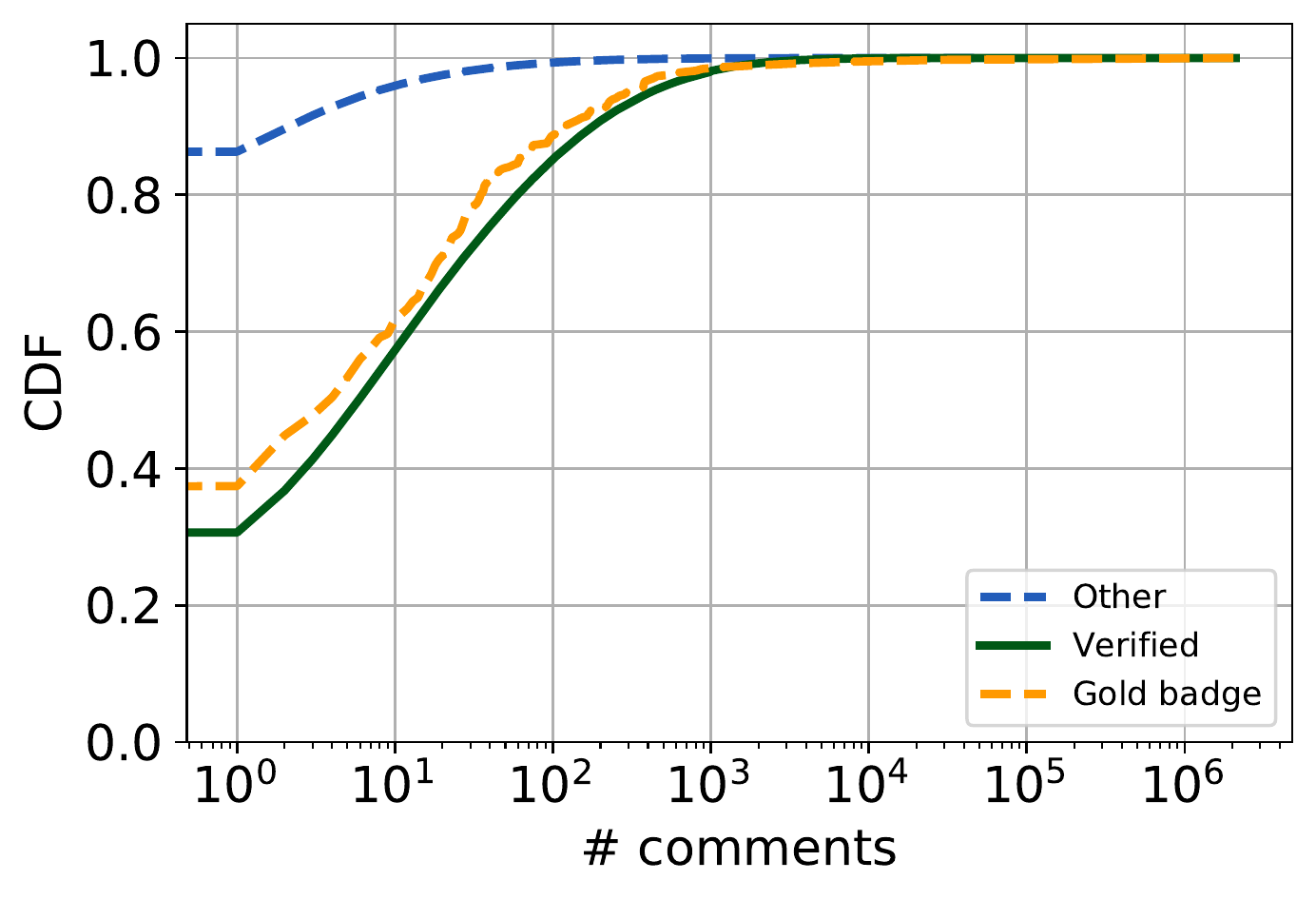}\label{fig:verified_comments_cdf}}
    \caption{CDFs of the number of posts and comments of verified, gold badge, and other users. (Note log scale on x-axis).}
    \label{fig:post_comm_cdfs}
    \end{figure*}

\begin{figure*}[t!]
  \centering
	\subfigure[]{\includegraphics[width=0.8\columnwidth]{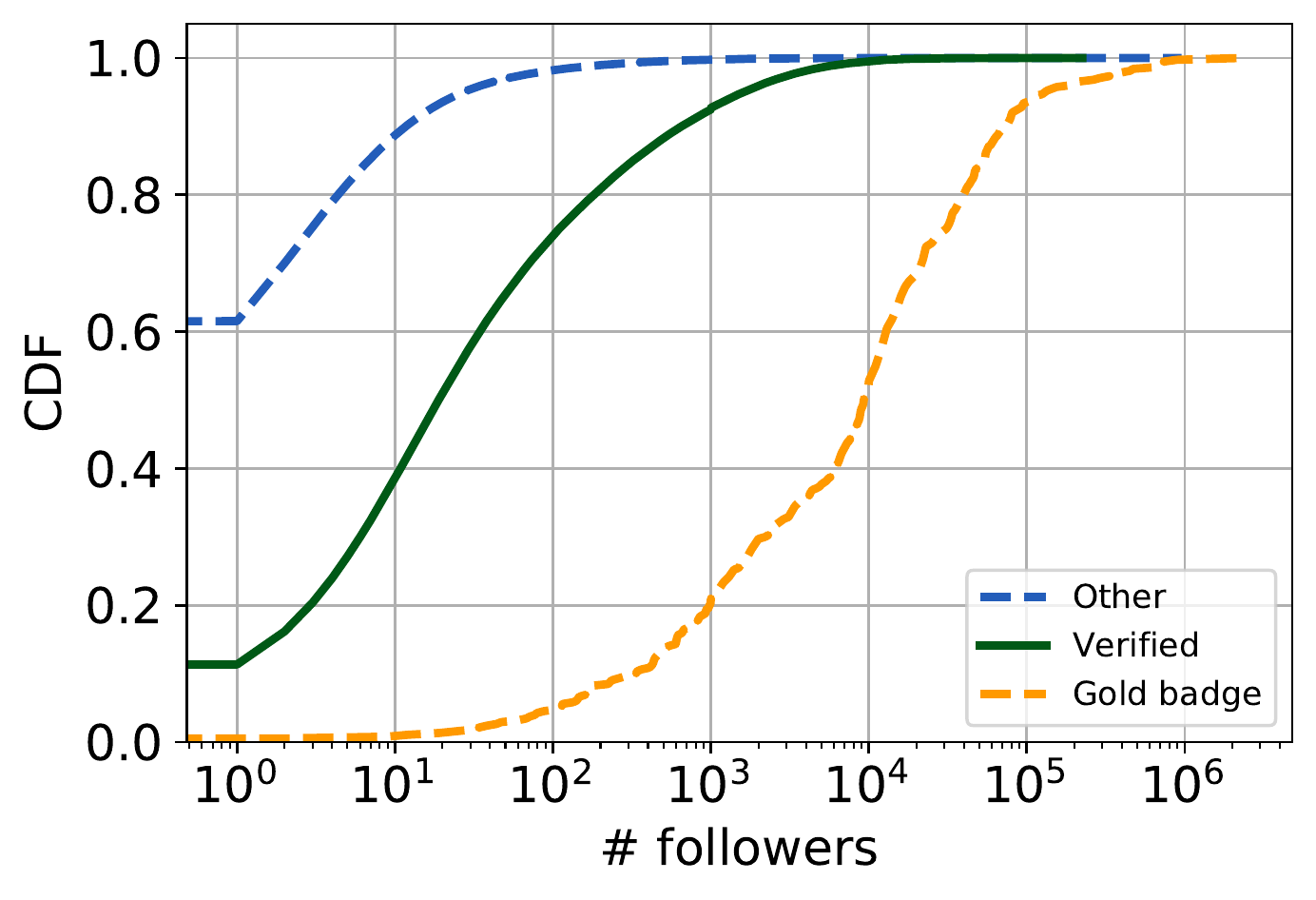}
  \label{fig:verified_follower_cdf}}
  ~~~
  \subfigure[]{\includegraphics[width=0.8\columnwidth]{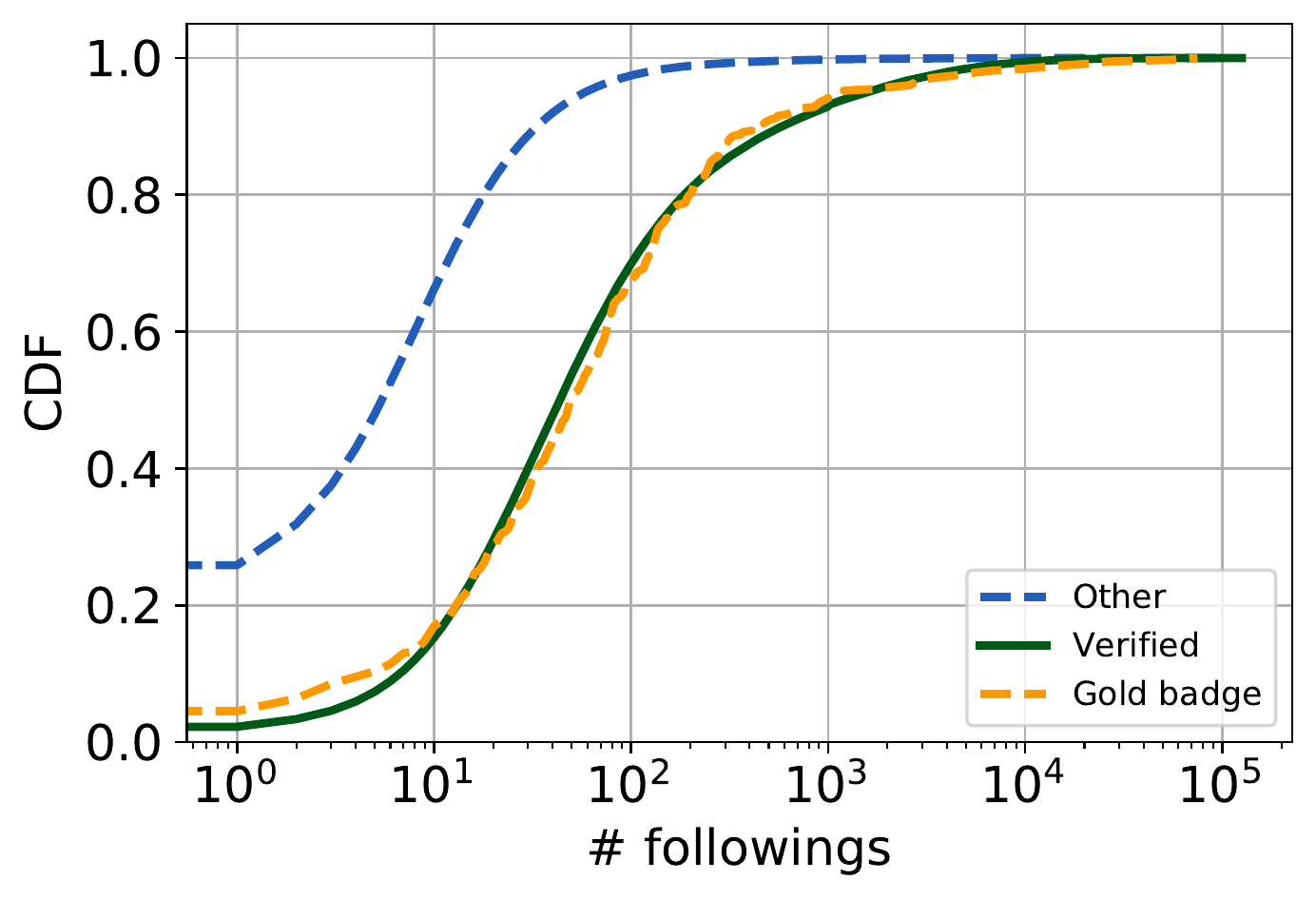}\label{fig:verified_following_cdf}}
  \caption{CDF of the number of following and followers of gold badge, verified, and other users. (Note log scale on x-axis).}
  \label{fig:foll_cdfs}
  \end{figure*}

\subsection{Gold badge users}
The user profile objects returned by the Parler API contain a ``verified'' field that corresponds to a boolean value.
All users with this value set to ``True'' have a gold badge and vice versa.
We assume that these users are actually the small set of truly ``verified'' users in the more widely adopted sense of the word, akin to ``blue check'' users on Twitter.
There are only 596 gold badge users on Parler, which is less than 1\% of the entire user count. This is in contrast to the red ``Verified'' badge tag (Badge \#0), which is awarded to users that undergo the identity check process.

From rudimentary exploratory analysis of the "Gold badge" verified users it appears that they are mostly a mixture of right wing celebrities, conservative politicians, conservative alternative media blogs, and conspiracy outlets. 
Some notable accounts are Rudy Giuliani and Enrique Tarrio, the recently arrested head of the Proud Boys~\cite{proudboys}. 
Of the 596 accounts we found with the gold badge, 51 of them had either "Trump" or "maga" in their bios.

For the remainder of the analysis we take into account gold badge users, users who are verified and do not have a gold badge, and users who are not verified and do not have a gold badge (other). 
For example, Figure~\ref{fig:post_comm_cdfs} %
shows post and comment activity split by these categorizations.
We see that both gold badge and verified users are overall more active than the ``other'' users.

There are only 4 users who are both gold badge and private.
These users are ``ScottMason,'' ``userfeedback,'' ``AFPhq'' (Americans for Prosperity), and ``govgaryjohnson'' (Governor Gary Johnson).

\subsection{Followers/Followings}

There are two numbers related to the underlying social network structure available in the user profile objects:
A user's followers corresponds to how many individuals are currently following that user, whereas their followings corresponds to how many users they follow.
Figure~\ref{fig:verified_follower_cdf} shows the cumulative distribution function (CDF) of the followers per user split by badge type, while Figure~\ref{fig:verified_following_cdf} shows the same for the followings of each user.
First we note that standard users are less popular, as they have fewer followers; see Figure~\ref{fig:verified_follower_cdf}.
Gold badge users on the other hand have a much larger number of followers.
We see that about 40\% of the typical users have more than a single follower, whereas about 40\% of gold badge users have more than 10,000 followers; verified users fall somewhere in the middle. 
As seen in Figure~\ref{fig:verified_following_cdf}, typical users  also follow a smaller number of accounts compared to gold and verified users. 

\subsection{User account creation}

\begin{figure}[t!]
    \centering
    \includegraphics[width=\columnwidth]{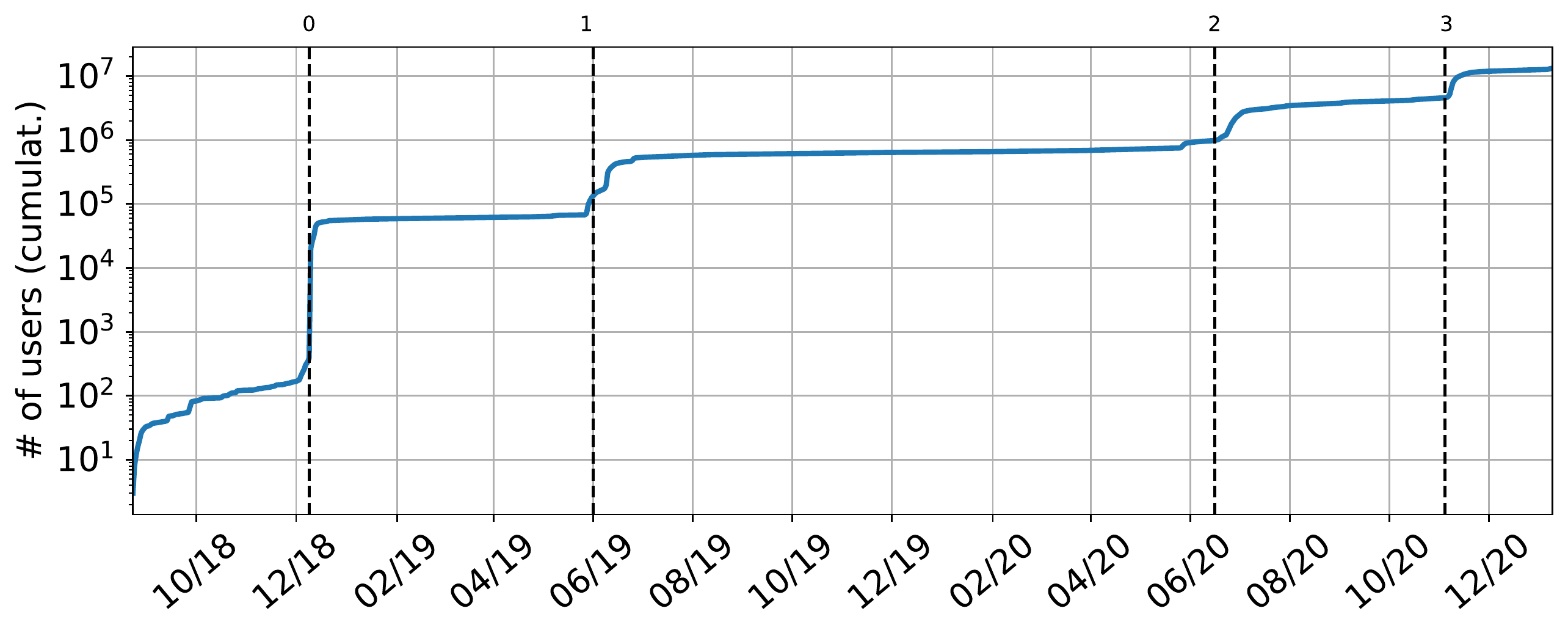}
    \caption{Cumulative number of users joining daily. (Note log scale on y-axis). Table~\ref{tab:events} reports the events annotated in the figure.}
    \label{fig:users_joined_cumulative}
    \end{figure}

\begin{figure}[t!]
    \centering
    \subfigure[November 2020]{\includegraphics[width=\columnwidth]{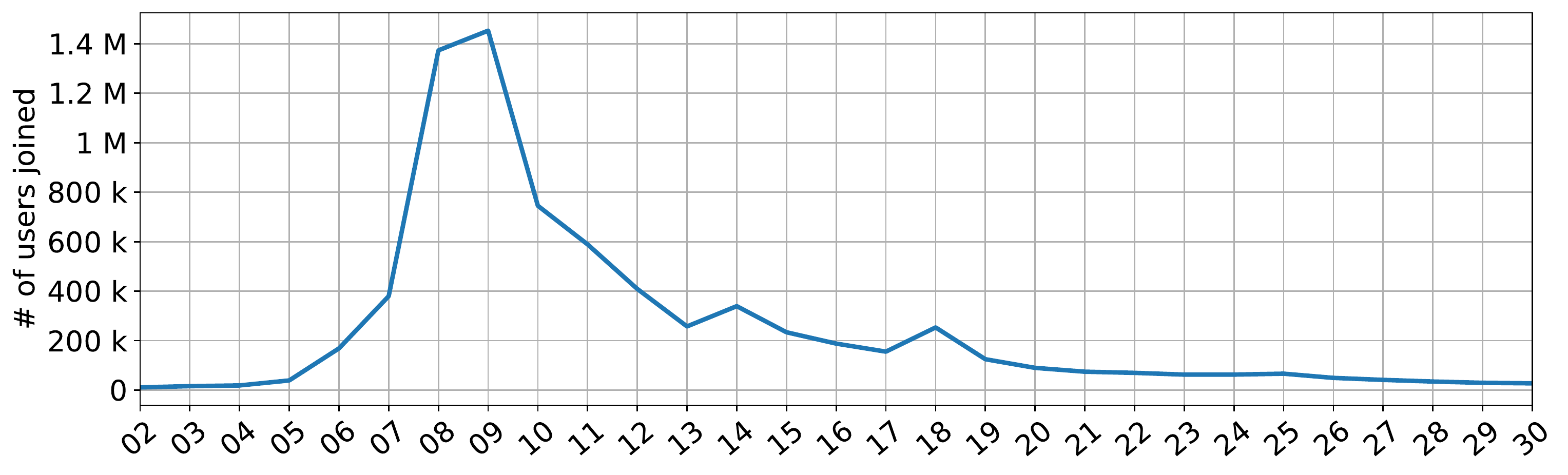}
    \label{fig:users_joined_nov_day}}
  \subfigure[January 2021]{\includegraphics[width=\columnwidth]{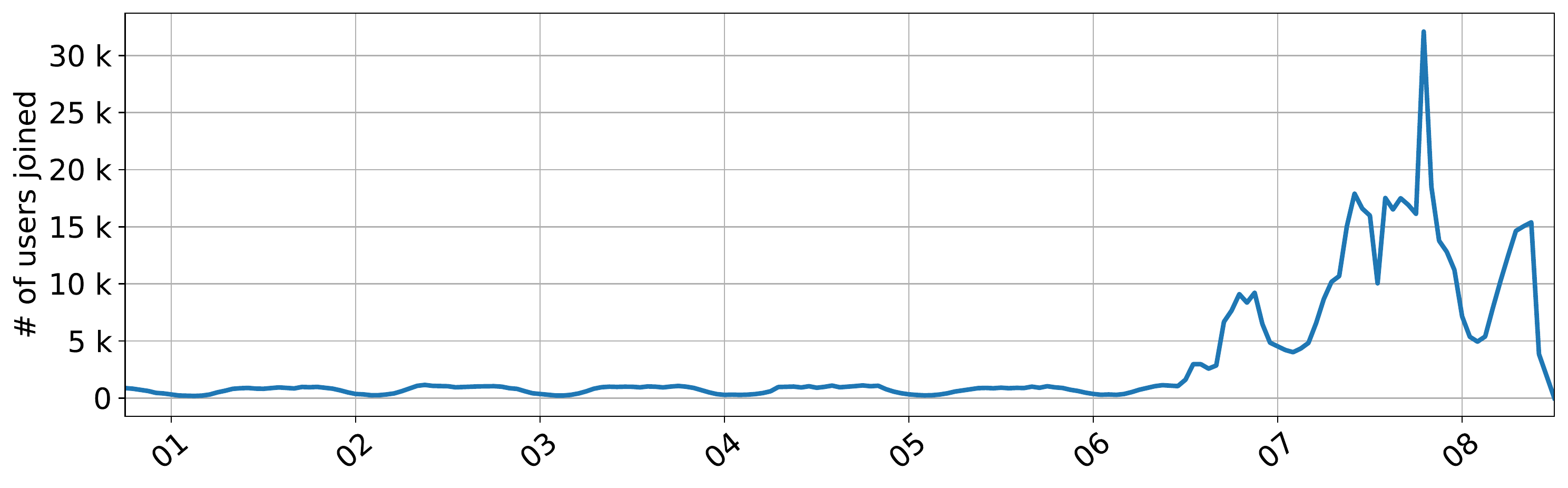}
  \label{fig:users_joined_jan}}
  \caption{Number of users joining per day in November 2020 (the month of the US Elections) and January 2021 (the month of the January 6 insurrection).}
  \end{figure}

The number of users on Parler grew throughout the course of the platform's lifetime.
We notice several key events correlated with periods of user growth.
Parler originally launched in August 2018. 
Figure~\ref{fig:users_joined_cumulative} plots cumulative users growth since Parler went live.
Parler saw its first major growth in number of users in December 2018, reportedly because conservative activist Candace Owens tweeted about it~\cite{candance}.
The second large new user event occurred in June 2019 when Parler reported that a large number of accounts from Saudi Arabia joined~\cite{reuters_saudi_arabia}.
In 2020 there were two large events of new users.
The first occurred in June 2020, where on June 16th 2020 conservative commentator Dan Bongino announced he had purchased an ownership stake in the platform~\cite{bongino}.
At the same time, Parler also received a second endorsement from Brad Parscale, the social media campaign manager for Trump's 2016 campaign. 
The last major user growth event in 2020 occurred around the time of the United States 2020 election and some cite~\cite{election} this growth as a result of Twitter's continuous fact-checking of Donald Trump's tweets.
As we show in Figure~\ref{fig:users_joined_nov_day}, a substantial number of new accounts were created during November 2020, while the outcome of the US 2020 Presidential Election was determined.
In addition, we saw additional substantial user growth in January 2021, as shown in Figure~\ref{fig:users_joined_jan}, especially in the days after the Capitol insurrection.

\begin{figure}[t!]
  \centering
  \includegraphics[width=\columnwidth]{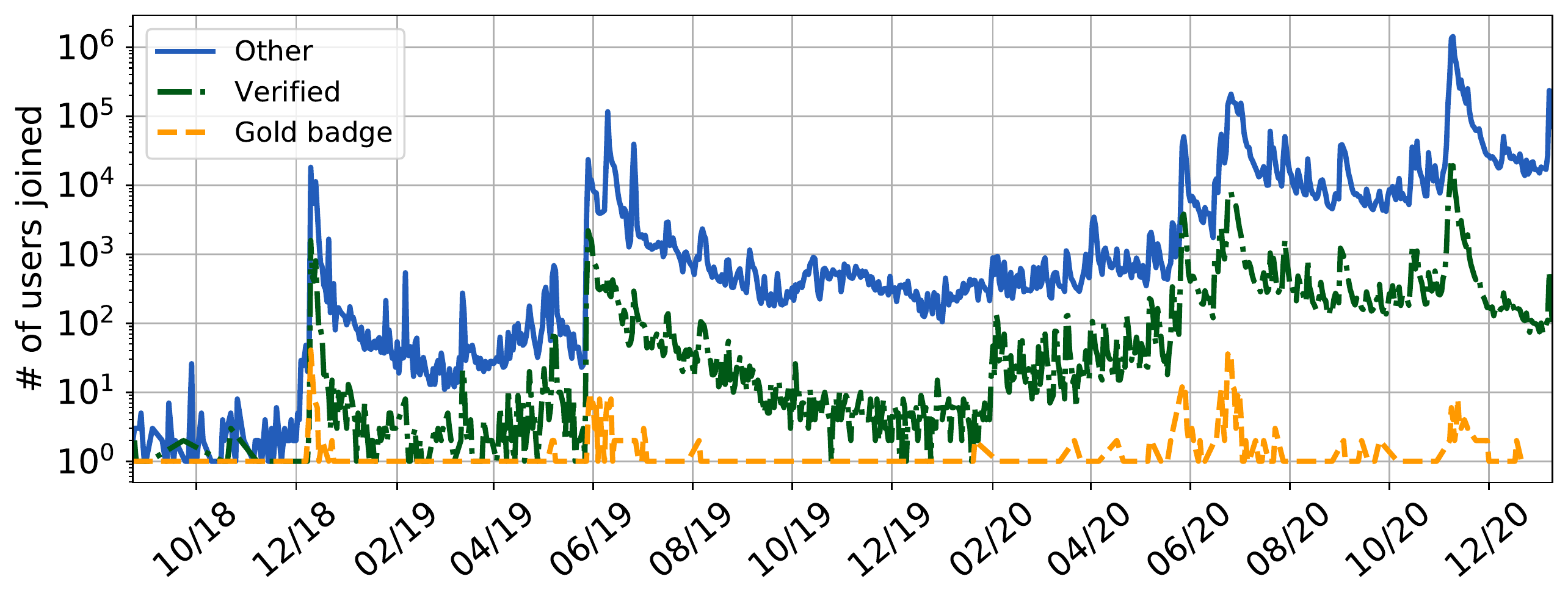}
  \caption{Number of users joining daily split by gold badge, verified, and other users. (Note log scale on y-axis.)}
  \label{fig:users_joined_verified}
  \end{figure}

Finally, Figure~\ref{fig:users_joined_verified} shows account creations for users that have a ``gold'' badge, ``verified'' badge, and ``other'' users that do not have either badge. 
We observe that throughout the course of time, Parler attracts new users that become verified and gold users.

\section{Content Analysis}
\label{sec:content_analysis}

We now analyze the content posted by Parler users, focusing on activity volume, voting, hashtags used, and URLs shared on the platform.

\begin{table}[t]
\centering
\small
\setlength{\tabcolsep}{2.5pt}
\begin{tabular}{@{}r>{\raggedright\arraybackslash}p{5.5cm}l@{}}
\toprule
{\textbf{\begin{tabular}[t]{@{}c@{}}Event \\ ID\end{tabular}}} & \multicolumn{1}{c}{\textbf{Description}}                                                            & \multicolumn{1}{c}{\textbf{Date}} \\ \midrule
0                                                                                & Candance Owens tweets about Parler \cite{candance}.                              & 2018-12-09                        \\
1                                                                                & Large amount of users from Saudi Arabia join Parler~\cite{reuters_saudi_arabia}.  & 2019-06-01                        \\
2                                                                                & Dan Bongino announces purchase of ownership stake on Parler~\cite{bongino}. & 2020-06-16                        \\
3                                                                                & 2020 US Presidential Election \cite{election}.                                                & 2020-11-04                        \\ \bottomrule
\end{tabular}%
\caption{Events depicted in Figures~\ref{fig:users_joined_cumulative} and~\ref{fig:counts_per_week}.}
\label{tab:events}
\end{table}

\begin{figure}[t!]
\centering
\includegraphics[width=\columnwidth]{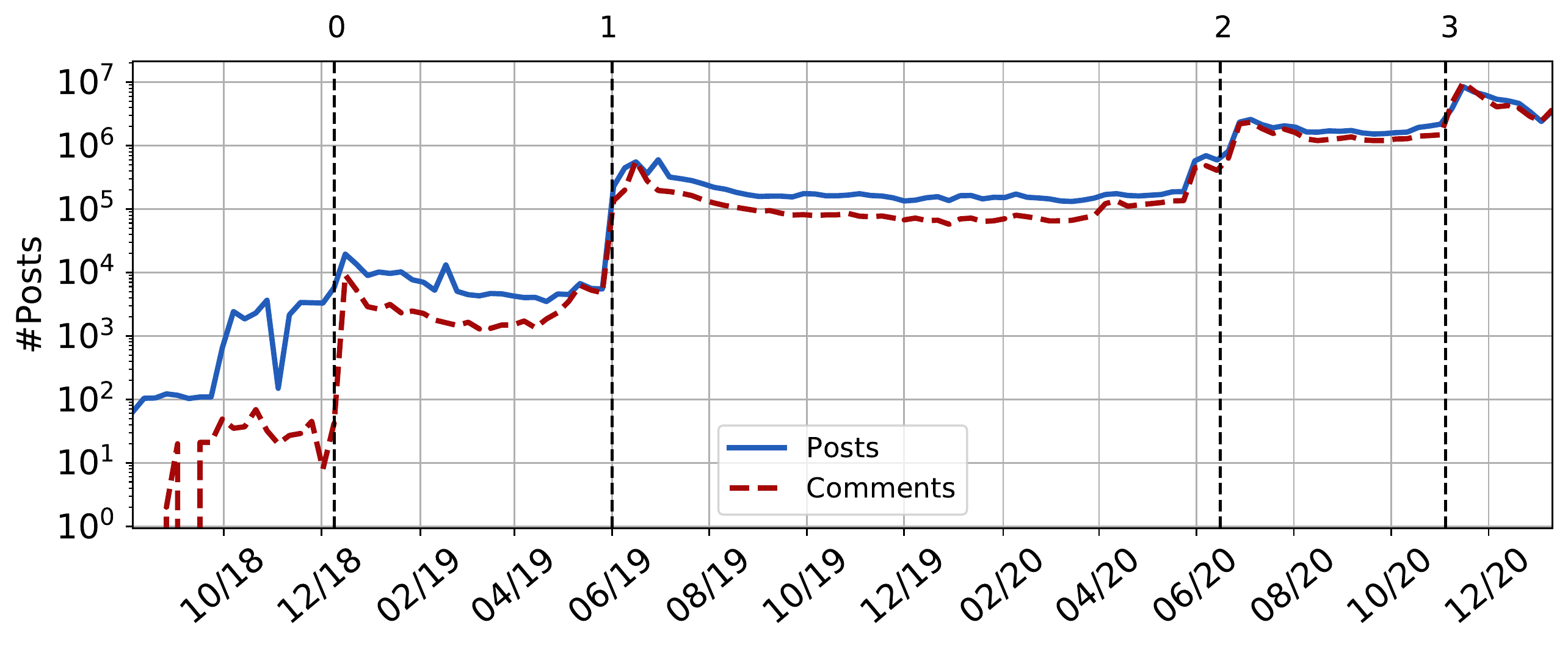}
\caption{Number of posts per week. (Note log scale on y-axis). Table~\ref{tab:events} reports the events annotated in the figure.}
\label{fig:counts_per_week}
\end{figure}

\begin{figure*}[t!]
\centering
\subfigure[Posts]{\includegraphics[width=0.75\columnwidth]{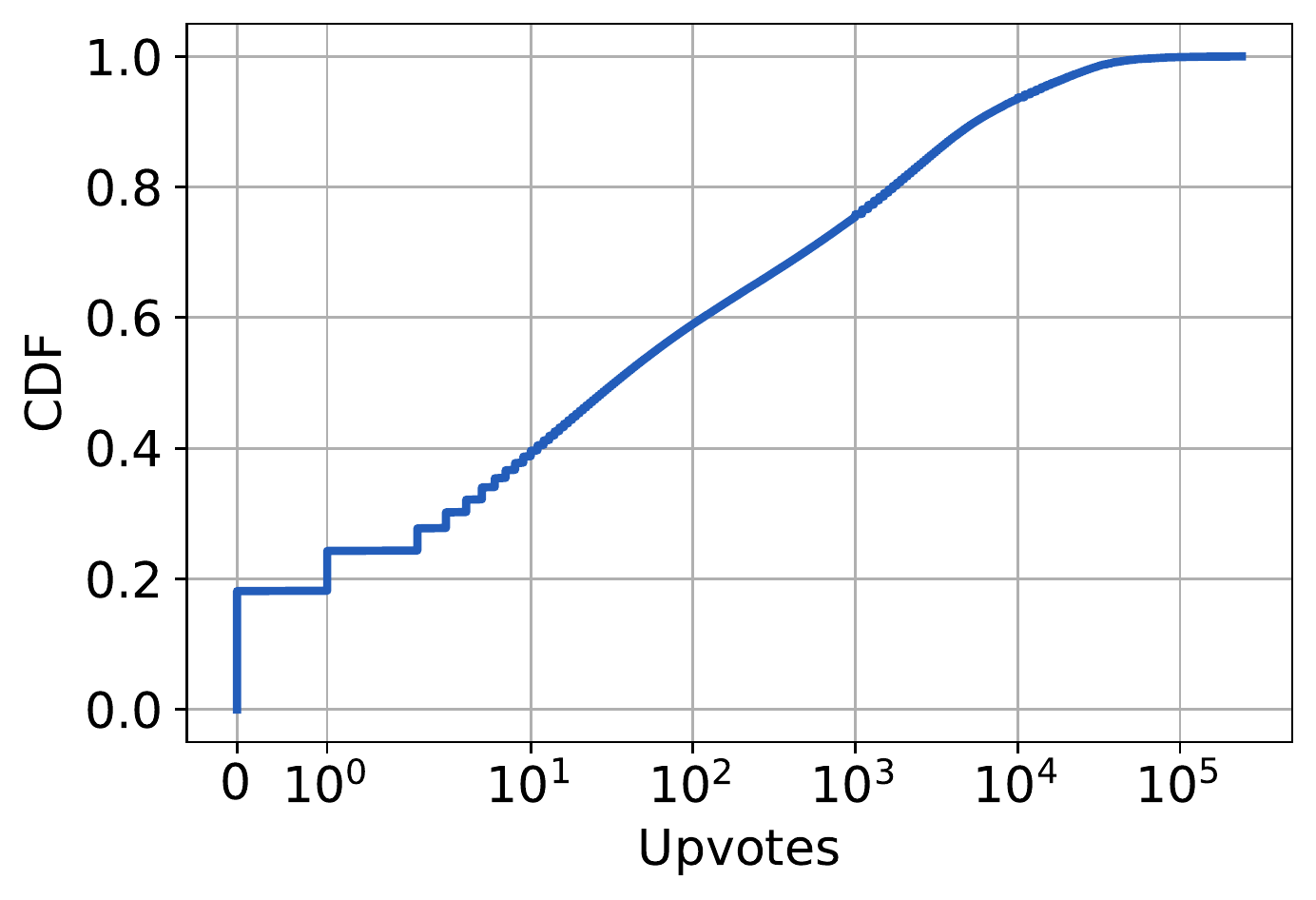}\label{fig:cdf_upvotes_per_sub}}
~~~
\subfigure[Comments]{\includegraphics[width=0.75\columnwidth]{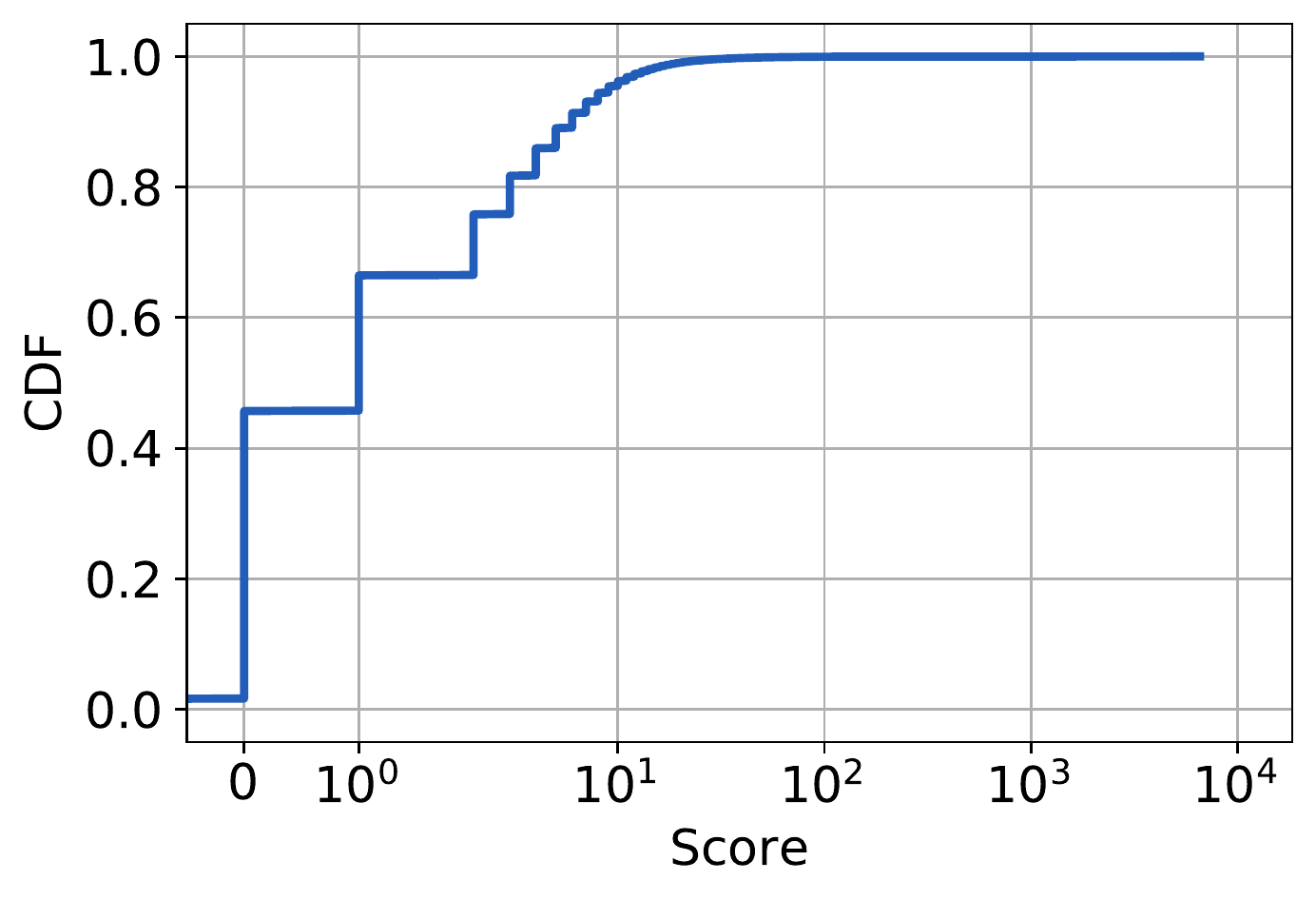}\label{fig:cdf_score_per_comment}}
\caption{CDFs of the number of upvotes on posts and scores (upvotes minus downvotes) on comments. (Note log scale on x-axis).} %
\label{fig:cdf_score}
\end{figure*}

\subsection{Activity Volume}
We begin our analysis by looking at the volume of posts and comments over time.
Figure~\ref{fig:counts_per_week} plots the weekly number of posts and comments in our dataset. %
We observe that the shape of curves is similar to that in Figure~\ref{fig:users_joined_cumulative}, i.e., there are spikes in post/comment activity at the same dates where there is an influx of new users due to external events.
Parler was a relatively small platform between August 2018 and June 2019, with less than 10K posts and comments per week.
Then, by June 2019, there is a substantial increase in the volume of posts and comments, with approximately 100K posts and comments per week. 
This coincides with a large-scale migration of Twitter users originating from Saudi Arabia, who joined Parler due to Twitter's ``censorship''~\cite{reuters_saudi_arabia}.
The volume of posts and comments remain relatively stable between June 2019 and June 2020, while, in mid-2020, there is another large increase in posts and comments, with 1M posts/comments per week. 
This coincides with when Twitter started flagging President's Trump tweets related to the George Floyd Protests, which prompted Parler to launch a campaign called ``Twexit,'' nudging users to quit Twitter and join Parler~\cite{twexit}.
Finally, by the end of our dataset in late 2020, another substantial increase in posts/comments coincides with a sudden interest in the platform after Donald Trump's defeat in the 2020 US Presidential Election.
Note that the number of posts per user is nearly constant except when there is a large influx of new users.

\subsection{Voting}
As mentioned in Section~\ref{sec:background}, posts on Parler can be upvoted, while comments can be upvoted and downvoted, thus yielding a score (sum of upvotes minus the sum of downvotes).
Figure~\ref{fig:cdf_score} shows the CDFs of the upvotes and score for posts and comments.
We find that 18\% of the posts do not receive any upvotes, and 61\% of posts receive at least 10 upvotes.
Looking at the scores for comments (see Figure~\ref{fig:cdf_score_per_comment}), we observe that comments rarely have a negative score (only 1.6\%), while 44\% of comments have a score equal to zero, and the rest  have positive scores.
Overall, our results indicate that a substantial amount of content posted on Parler is viewed positively by its users. %

\subsection{Hashtags}
Next, we focus on the prevalence and popularity of hashtags on Parler. 
We find that only a small percentage of posts/comments include hashtags: 2.9\% and 3.4\% of all posts and comments, respectively.
We then analyze the most popular hashtags, as they can provide an indication of users' interests.
Table~\ref{tab:top_hashtags} reports the top 20 hashtags in posts and comments.
Among the most popular hashtags in posts (left side of the table), we find \#trump2020, \#maga, and \#trump, which suggests that many of Parler's users are Trump supporters and discuss the 2020 US elections.
We also find hashtags referring to conspiracy theories, such as \#wwg1wga, \#qanon, and \#thegreatawakening, which refer to the QAnon conspiracy theory~\cite{originsqanon,papasavva2020qoincidence}.\footnote{Where We Go One We Go All (WWG1WGA) is a popular QAnon motto.}
Furthermore, we find several hashtags that are related to the alleged election fraud that Trump and his supporters claimed occured during the 2020 US Elections (e.g., \#stopthesteal, \#voterfraud, and \#electionfraud).

In comments (see right side of Table~\ref{tab:top_hashtags}), we observe that the most popular hashtag is \#parlerconcierge.
A manual examination of a sample of the posts suggests that this hashtag is used by Parler users to welcome new users (e.g., when a new user makes their first post, another user replies with a comment including this hashtag).
Similar to posts, we find thematic use of hashtags showing support for Donald Trump and conspiracy theories like QAnon.

\begin{table}[t]
\centering
\small
\setlength{\tabcolsep}{2.5pt}
\begin{tabular}{@{}lrlr@{}}
\toprule
\textbf{Hashtag}  & \multicolumn{1}{r}{\textbf{\#Posts}} & \textbf{Hashtag}   & \textbf{\#Comments} \\ \midrule
trump2020         & \multicolumn{1}{r|}{347,799}         & parlerconcierge    & 962,207                                 \\
maga              & \multicolumn{1}{r|}{271,379}         & trump2020          & 181,219                                 \\
stopthesteal           & \multicolumn{1}{r|}{200,059}         & newuser               & 157,186                                 \\
parler      & \multicolumn{1}{r|}{187,363}         & maga       & 148,655                                  \\
wwg1wga            & \multicolumn{1}{r|}{176,150}         & truefreespeech            & 147,769                                  \\
trump             & \multicolumn{1}{r|}{168,649}          & stopthesteal             & 93,927                                  \\
kag             & \multicolumn{1}{r|}{117,894}          & wwg1wga                & 52,687                                  \\
qanon               & \multicolumn{1}{r|}{117,134}          & parler              & 45,211                                  \\
freedom         & \multicolumn{1}{r|}{108,794}          & kag           & 43,396                                  \\
parlerksa               & \multicolumn{1}{r|}{97,004}          & trump                & 39,659                                  \\
newuser          & \multicolumn{1}{r|}{87,263}          & maga2020          & 31,055                                  \\
news           & \multicolumn{1}{r|}{86,771}          & usa                  & 28,046                                  \\
usa      & \multicolumn{1}{r|}{84,271}          & obamagate        & 26,250                                  \\
trumptrain         & \multicolumn{1}{r|}{82,893}          & 1          & 22,850                                  \\
thegreatawakening        & \multicolumn{1}{r|}{82,710}          & wethepeople            & 22,236                                  \\
meme           & \multicolumn{1}{r|}{82,440}          & fightback & 21,954                                  \\
electionfraud     & \multicolumn{1}{r|}{80,457}          & blm                & 19,979                                  \\
maga2020        & \multicolumn{1}{r|}{79,046}          & qanon              & 19,758                                  \\
voterfraud              & \multicolumn{1}{r|}{78,793}          & trump2020landslide       & 19,179                                  \\
americafirst & \multicolumn{1}{r|}{75,764}          & americafirst                 & 19,012                                  \\ \bottomrule
\end{tabular}%
\caption{Top 20 hashtags in posts and comments.}
\label{tab:top_hashtags}
\end{table}

\subsection{URLs}
Finally, we focus on URLs shared by Parler users: 15.7\% and 7.9\% of all posts and comments, respectively, include at least one URL.
Table~\ref{tab:top-domains} reports the top 20 domains in the shared URLs. %
Among the most popular domains, we find Parler itself, YouTube, image hosting sites like Imgur, links to mainstream social media platforms like Twitter, Facebook, and Instagram, as well as news sources like Breitbart and New York Post.

Overall, our URL analysis suggests that Parler users are sharing a mixture of both mainstream and alternative content on the Web.
For instance, they are sharing YouTube URLs (mainstream) as well as Bitchute URLs, a ``free speech'' oriented YouTube alternative~\cite{benHorneBitchute}. 
The same applies with news sources: Parler users are sharing both alternative news sources (e.g., Breitbart) and mainstream ones (New York Post, a conservative-leaning outlet), with the alternative news sources being more popular in general.  

\begin{table}[t]
\centering
\setlength{\tabcolsep}{1.5pt}
\small
\begin{tabular}{lrlr}
\toprule
\textbf{Domain} & \textbf{\#Posts} &     \textbf{Domain} & \textbf{\#Comments} \\
\midrule
parler.com &  \multicolumn{1}{r|}{5,017,486} &  parler.com &  2,488,718 \\
youtu.be &  \multicolumn{1}{r|}{1,275,127} &  youtube.com &  1,767,928 \\
youtube.com &  \multicolumn{1}{r|}{827,145} &  giphy.com &  1,314,282 \\
twitter.com &  \multicolumn{1}{r|}{773,041} &  bit.ly &  872,611 \\
facebook.com &  \multicolumn{1}{r|}{493,804} &  youtu.be &  449,666 \\
thegatewaypundit.com\hspace*{-0.15cm} &  \multicolumn{1}{r|}{478,982} &  imgur.com &  163,381 \\
imgur.com &  \multicolumn{1}{r|}{353,184} &  par.pw &  50,779 \\
breitbart.com &  \multicolumn{1}{r|}{345,700} &  twitter.com &  35,035 \\
foxnews.com &  \multicolumn{1}{r|}{336,390} &  tenor.com &  32,922 \\
theepochtimes.com &  \multicolumn{1}{r|}{236,278} &  bitchute.com &  30,751 \\
giphy.com &  \multicolumn{1}{r|}{87,344} &  facebook.com &  27,460 \\
instagram.com &  \multicolumn{1}{r|}{162,769} &  rumble.com &  19,919 \\
rumble.com &  \multicolumn{1}{r|}{142,495} &  thegatewaypundit.com\hspace*{-0.4cm} &  12,747 \\
westernjournal.com &  \multicolumn{1}{r|}{99,271} &  google.com &  12,249 \\
t.co &  \multicolumn{1}{r|}{84,633} &  whitehouse.gov &  12,002 \\
nypost.com &  \multicolumn{1}{r|}{84,288} &  blogspot.com &  11,575 \\
par.pw &  \multicolumn{1}{r|}{78,473} &  gmail.com &  9,267 \\
ept.ms &  \multicolumn{1}{r|}{77,069} &  wordpress.com &  9,181 \\
bitchute.com &  \multicolumn{1}{r|}{73,970} &  amazon.com &  8,886 \\
townhall.com &  \multicolumn{1}{r|}{72,781} &  foxnews.com &  7,987 \\
\bottomrule
\end{tabular}
\caption{Top domains on Parler. }
\label{tab:top-domains}
\end{table}

\section{Related Work}

In this section, we review related work.

\descr{Datasets.} %
~\cite{brena2019news} present a data collection pipeline and a dataset with news articles along with their associated sharing activity on Twitter.
\cite{fair2019shouting} release a dataset of 37M posts, 24.5M comments, and 819K user profiles collected from Gab. %
\cite{papasavva2020raiders} present an annotated dataset with 3.3M threads and 134.5M posts from the Politically Incorrect board (/pol/) of the imageboard forum 4chan, posted over a period of almost 3.5 years (June 2016--November 2019).
~\cite{baumgartner2020pushshift} present a large-scale dataset from Reddit that includes 651M submissions and 5.6B comments posted between June 2005 and April 2019.
\cite{garimella2018whatapp} present a methodology for collecting large-scale data from WhatsApp public groups and release an anonymized version of the collected data.
They scrape data from 200 public groups and obtain 454K messages from 45K users.
Finally,~\cite{founta2018crowdsourcing} use crowdsourcing to label a dataset of 80K tweets as normal, spam, abusive, or hateful.
More specifically, they release the tweet IDs (not the actual tweet) along with the majority label received from the crowd-workers.  

\descr{Fringe Communities.} Over the past few years, a number of research papers have provided data-driven analyses of fringe, alt- and far-right online communities, such as 4chan~\cite{hine2017kek,bernstein20114chan,tuters2019they,pettisambiguity}, Gab~\cite{zannettou2018gab}, Voat~\cite{papasavva2020qoincidence}, The\_Donald and other hateful subreddits~\cite{flores2018mobilizing,mittos2020and}, etc.
Prior work has also analyzed their impact on the wider web, e.g., with respect to disinformation~\cite{zannettou2017web}, hateful memes~\cite{zannettou2018origins}, and doxing~\cite{snyder2017fifteen}.

\section{Conclusion}\label{sec:conclusion}
This paper presented our Parler dataset, along with a general characterization. 
We collected and released user information for 13.25M users that joined the platform between 2018 and 2020, as well as a sample of 183M posts by 4M users.  

Our preliminary analysis shows that Parler attracts the interest of conservatives, Trump supporters, religious, and patriot individuals.
Also, the data reveals that Parler experienced large influxes of new users in close temporal proximity with real-world events related to online censorship on mainstream platforms like Twitter, as well as events related to US politics.
Additionally, our dataset sheds light into the content that is disseminated on Parler; for instance, Parler users share content related to US politics, content that show support to Donald Trump and his efforts during the 2020 US elections, and content related to conspiracy theories like the QAnon conspiracy theory.

Overall, Parler is an emerging alternative platform that needs to be considered by the research community that focuses on understanding emerging socio-technical issues (e.g., online radicalization, conspiracy theories, or extremist content) that exist on the Web and are related to US politics.
To this end, we are confident that our dataset will pave the way to motivate and assist researchers in studying and understanding extreme platforms like Parler, especially at a crucial point of US and World history.

At the time of writing, Parler is being taken down by its hosting provider, Amazon AWS.
It is unclear when and how the service will come back, which potentially makes the snapshot provided in this paper even more useful to the research community.

\small
\bibliographystyle{abbrv}
\bibliography{references}

\end{document}